\documentclass[epjCONF]{svjour}
\usepackage{graphics}
\usepackage[varg]{txfonts} 
\usepackage[latin1]{inputenc}

 \newcommand\beq{\begin{equation}}
 
 \newcommand\eeq{\end{equation}}
 \newcommand\beqn{\begin{eqnarray}}
 \newcommand\eeqn{\end{eqnarray}}
 \newcommand\nn{\nonumber}

\def\MeV{\,\mbox{MeV}}
\def\GeV{\,\mbox{GeV}}

\session-title{International Conference on New Frontiers in Physics}
\begin{document}
\title{Incurable Adler relation for soft neutrino interactions}
\author{B.Z. Kopeliovich\thanks{\email{boris.kopeliovich@usm.cl}} \and I.K. Potashnikova\thanks{\email{irina.potashnikova@usm.cl}} \and Iv\'an Schmidt\thanks{\email{ivan.schmidt@usm.cl}}
\and M.~Siddikov\thanks{\email{marat.siddikov@usm.cl}}}
\institute{Departamento de F\'{\i}sica, Universidad T\'ecnica Federico Santa Mar\'{\i}a,\\
Centro de Estudios
Subat\'omicos, and Centro Cient\'ifico-Tecnol\'ogico de Valpara\'iso,\\
Casilla 110-V, Valpara\'iso, Chile}
\abstract{
The Adler relation (AR), which bridges soft interactions of neutrinos and pions,
might look as a manifestation of pion dominance. However neutrino cannot fluctuate to a pion
because of lepton current conservation, instead it interacts via much heavier
hadronic components. This fact leads to nontrivial relations between
interaction amplitudes of different hadronic species, in particular, it links diagonal and 
off-diagonal diffractive interactions of pions. Absorptive corrections
break these relations making the AR impossible to hold universally, i.e. for any target and at any energy.
We predict a dramatic breakdown of the AR for coherent neutrino-production of pions
on nuclei at all energies.} 

\maketitle

\section{Introduction}
\label{intro}

In 1964 Stephen Adler derived the eminent relation \cite{adler} between the cross sections of forward neutrino-to-lepton scattering and pion cross section on the same hadronic target with the same final hadronic state.
Although it is tempting to interpret such a relation as pion pole dominance, it was realized \cite{bell,p-s,km} that this is not correct. Neutrino cannot fluctuate into a pion, $\nu\to l+\pi$, because of the transversity of the lepton current. 
Such a transition is completely forbidden for neutral currents and suppressed by the lepton mass squared for charge current processes. The meaning of the Adler relation (AR) is more delicate. Hadronic fluctuations of a neutrino are much heavier than the pion (even heavier than the $\rho$-meson), but miraculously the neutrino cross section
mediated by such heavy fluctuations turns out to be related to the pion interaction. The underlying dynamics of such 
a link is unknown, as well as  the mechanism of chiral symmetry breaking. It is based on the phenomenon
of partial conservation of axial current (PCAC).
We demonstrate here that strong absorptive corrections on nuclear targets cause a dramatic suppression of diffractive neutrino-production of pions compared with the expectations based on the AR.

\section{PCAC}
\label{pcac}

In the chiral limit of massless quarks both the vector and axial currents are conserved: 
\beqn
q_\mu V_\mu&=&q_\mu\,[\bar q(k')\,\gamma_\mu\,q(k)] = 0;\\ \nn
q_\mu A_\mu&=&q_\mu\,[\bar q(k')\,\gamma_5\gamma_\mu\,\vec\tau\,q(k)] = 0,
\label{10}
\eeqn
where $q_\mu=k^\prime_\mu-k_\mu$.

Hadrons acquire large masses via the mechanism of spontaneous 
symmetry breaking. The hadronic currents may be still conserved. 
For the vector current this is rather obvious:
\beq
q_\mu\,j_\mu^V=q_\mu\,\bar p(k')\,\gamma_\mu\,n(k) = (m_n-m_p)\bar p\,n = 0,
\label{20}
\eeq
up to the QED corrections.

Conservation of axial current looks more involved:
\beq
q_\mu\,j_\mu^A=q_\mu\,\bar p(k')\,\gamma_\mu\gamma_5\,n(k) = (m_n+m_p)\bar p\,\gamma_5\,n \neq 0.
\label{30}
\eeq

Nevertheless, in the general form,
\beq
j_\mu^A=\bar p(k')\left[g_A\,\gamma_\mu\gamma_5 - g_p\,q_\mu\gamma_5\right]n(k),
\label{40}
\eeq

the axial current can be conserved if
\beq
g_P(Q^2)=g_A(Q^2)\frac{2m_N}{Q^2}.
\label{50}
\eeq
The pole behavior here shows presence 
of a massless Goldstone particle \cite{goldstone}.
This proves the Goldstone theorem:  spontaneous breaking of chiral symmetry generates massless particles identified with pions.

\subsection{Goldberger-Treiman conspiracy}\label{conspiracy}

\begin{figure}[h!]
\centerline{
\resizebox{0.4\columnwidth}{!}{
\includegraphics{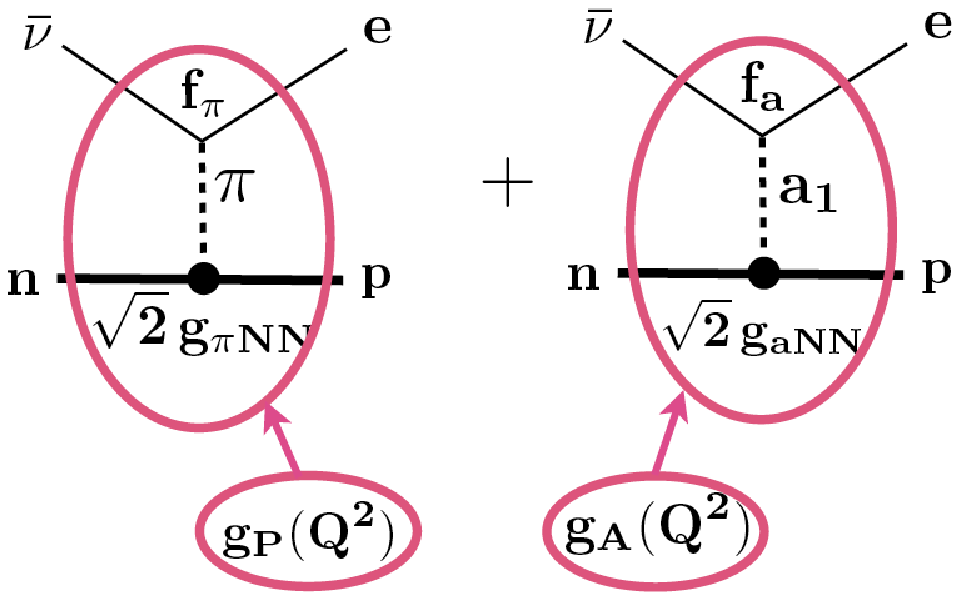} }\hspace{0.5cm}
\resizebox{0.5\columnwidth}{!}{
\includegraphics{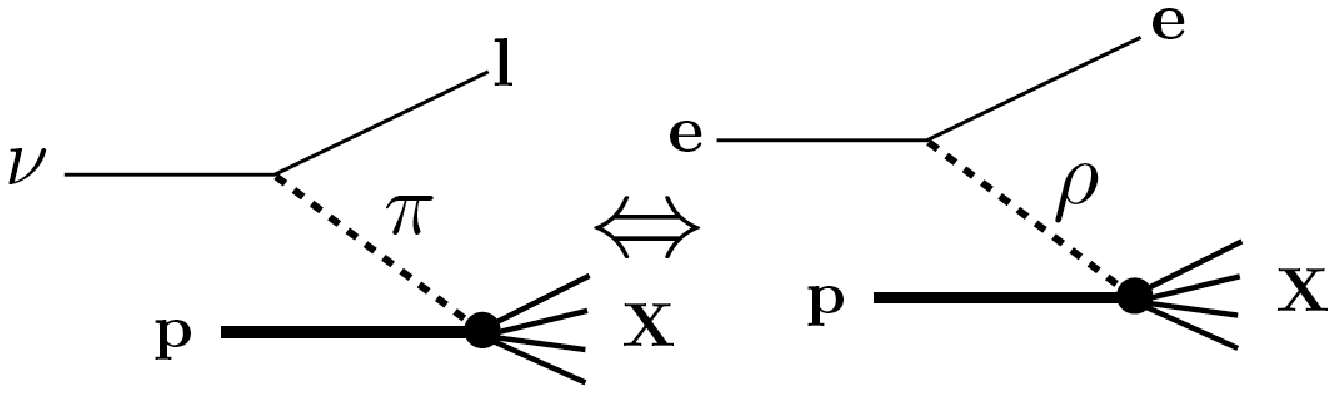}}}
\caption{{\it Left:} graphical representation for the effective pseudo-scalar and axial-vector couplings in Eq.~(\ref{40}).
{\it Right:} fake analogy between pion pole and $\rho$-meson dominance in production of hadronic states $X$ in lepton-proton interactions.} 
\label{fig:conspiracy}      
\end{figure}

PCAC leads to a miraculous relation between the quantities having very different origins.
\beq
\sqrt{2}m_N\,g_A(0)=f_\pi\,g_{\pi NN},
\label{60}
\eeq
where $f_\pi$ and $g_{\pi NN}$ are the the pion decay and pion-nucleon couplings, respectively, both known from data. 
It is tempting to interpret this 
Goldberger-Treiman relation \cite{treiman}  in
terms of pion pole dominance, like is illustrated  by the first term in the left panel of Fig.~\ref{fig:conspiracy}.
However, the pion pole does not contribute to $\beta$-decay, 
because the lepton current is conserved (up to the electron mass),
$\Gamma(\pi\to\bar\nu e)\ \propto\ m_e^2$.

The axial-vector formfactor  $g_A(Q^2)$ represents the contribution of heavy states, which  are related to the pion term via PCAC. John Bell called this relation between the axial constant and the pion pole Goldberger-Treiman conspiracy \cite{bell}.

\subsection{Hadronic properties of neutrinos}\label{AR}

Although the non-trivial Goldberger-Treiman relation is well confirmed by data for 
neutron decay and muon capture, the PCAC hypothesis should be
tested thoroughly in other processes.

The Fock components of a high-energy neutrino at low scale are dominated by the axial-vector hadronic fluctuations, since the vector term vanishes at  $Q^2\to0$ due to CVC.
The amplitude of the process $\nu+p\to l+X$, where $X$ is the final hadronic state, has the form,
\beq
M=\frac{G}{\sqrt{2}}\,l_\mu\left(V_\mu+A_\mu\right),
\label{70}
\eeq
where the conserved lepton current reads,
\beq
l_\mu=\bar l(k')\,\gamma_\mu(1+\gamma_5)\,\nu(k).
\label{80}
\eeq
At  $Q^2\to0$    the vector current contribution and the transverse part
of the axial term vanish, only  $\sigma^A_L$         survives, 
correspondingly, the matrix element squared has the factorized form,
\beq
\left|{\overline M}\right|^2=\frac{G^2}{2}\,L_{\mu\nu}\,A_{\mu\nu},
\label{90}
\eeq
where the lepton tensor $L_{\mu\nu}$ in the limit $Q^2\to0$ reads,
\beq
L_{\mu\nu}(Q^2\to0)=2\,\frac{E_\nu(E_\nu-\nu)}{\nu^2}\,q_\mu q_\nu.
\label{100}
\eeq
Here $E_\nu$ is the energy of the neutrino, and $\nu=E_\nu-E_l$ is the transferred energy in the target rest frame.
Remarkably, this tensor is proportional to $q_\mu q_\nu$, so one can apply the PCAC relation in the form \cite{km},
\beq 
q_\mu\, j^A_\mu = m_\pi^2\,\phi_\pi,
\label{110}
\eeq
which leads to the Adler relation (AR),
\beq
\left.\frac{d^2\sigma(\nu p\to l\,X)}{dQ^2\,d\nu}\right|_{Q^2=0} =
\frac{G^2}{2\pi^2}\,f_\pi^2\,\frac{E_\nu-\nu}{E_\nu\nu}\,\sigma(\pi p\to X).
\label{120}
\eeq

In analogy to the vector dominance model it is tempting to interpret the Adler relation as a manifestation of pion dominance, as is illustrated in Fig.~\ref{fig:conspiracy} (right panel).

However, neutrinos do not fluctuate to pions because of conservation of 
the lepton current, $q_\mu l_\mu=0$ in Eq.~(\ref{70}), so the pion intermediate state does not contribute, i.e. the above interpretation of the AR is not correct.

To understand the physics of the AR, let us single out the main singularities in the  dispersion relation for the axial current amplitude,
\beq
A_\mu(Q^2)=\frac{f_\pi\,q_\mu}{Q^2+m_\pi^2}\,T(\pi p\to X) + 
\frac{f_{a_1}}{Q^2+m_{a_1}^2}\,T_\mu(a_1 p\to X)\,+\,...
\label{130}
\eeq
The first term, the pion pole, does not contribute to the process due to conservation of the lepton current, $q_\mu l_\mu=0$. So the intermediate hadronic states must be heavier axial-vector hadrons, e.g. the $a_1$ meson, which is the chiral partner of the $\rho$-meson. It seems to be natural to expect $a_1$ dominance in the axial current, in analogy with the vector meson dominance. In this case one can restrict the dispersion relation for the axial current by the first two terms, explicitly shown in Eq.~(\ref{130}). Then these terms must be related in order to cancel in $q_\mu A_\mu$ and provide PCAC, eq.~(\ref{110}).

Such a nontrivial relation between the two terms in (\ref{130}) was challenged by Piketty and Stodolsky \cite{p-s}, who found that the  AR leads to the equality $\sigma_{diff}(\pi p\to a_1p)\approx \sigma_{el}(\pi p\to \pi p)$, which contradicts data by factor 20(!). 

\begin{figure}[h!]
\centerline{
\resizebox{0.4\columnwidth}{!}{
\includegraphics{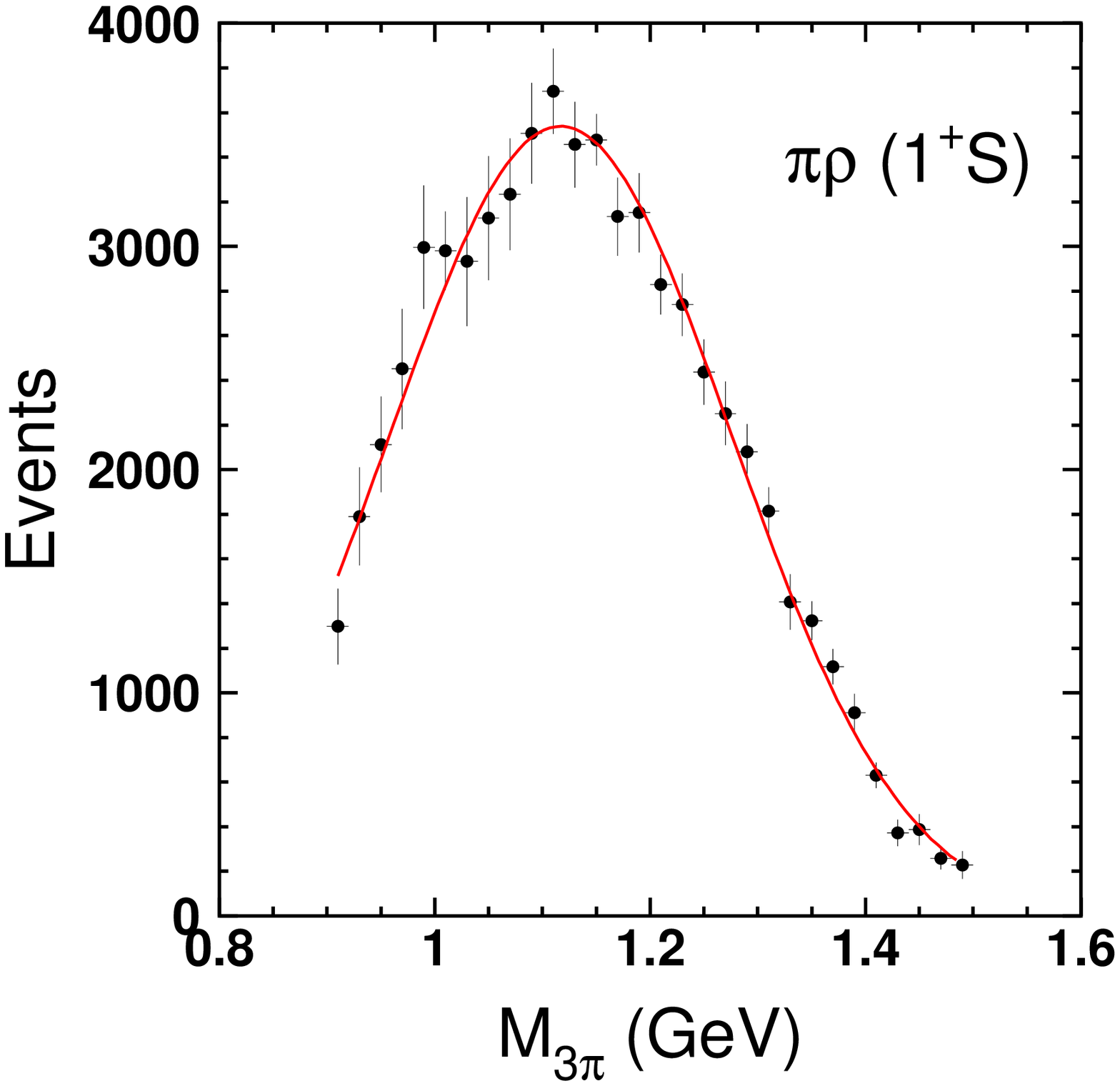} }\hspace{2cm}
\resizebox{0.35\columnwidth}{!}{
\includegraphics{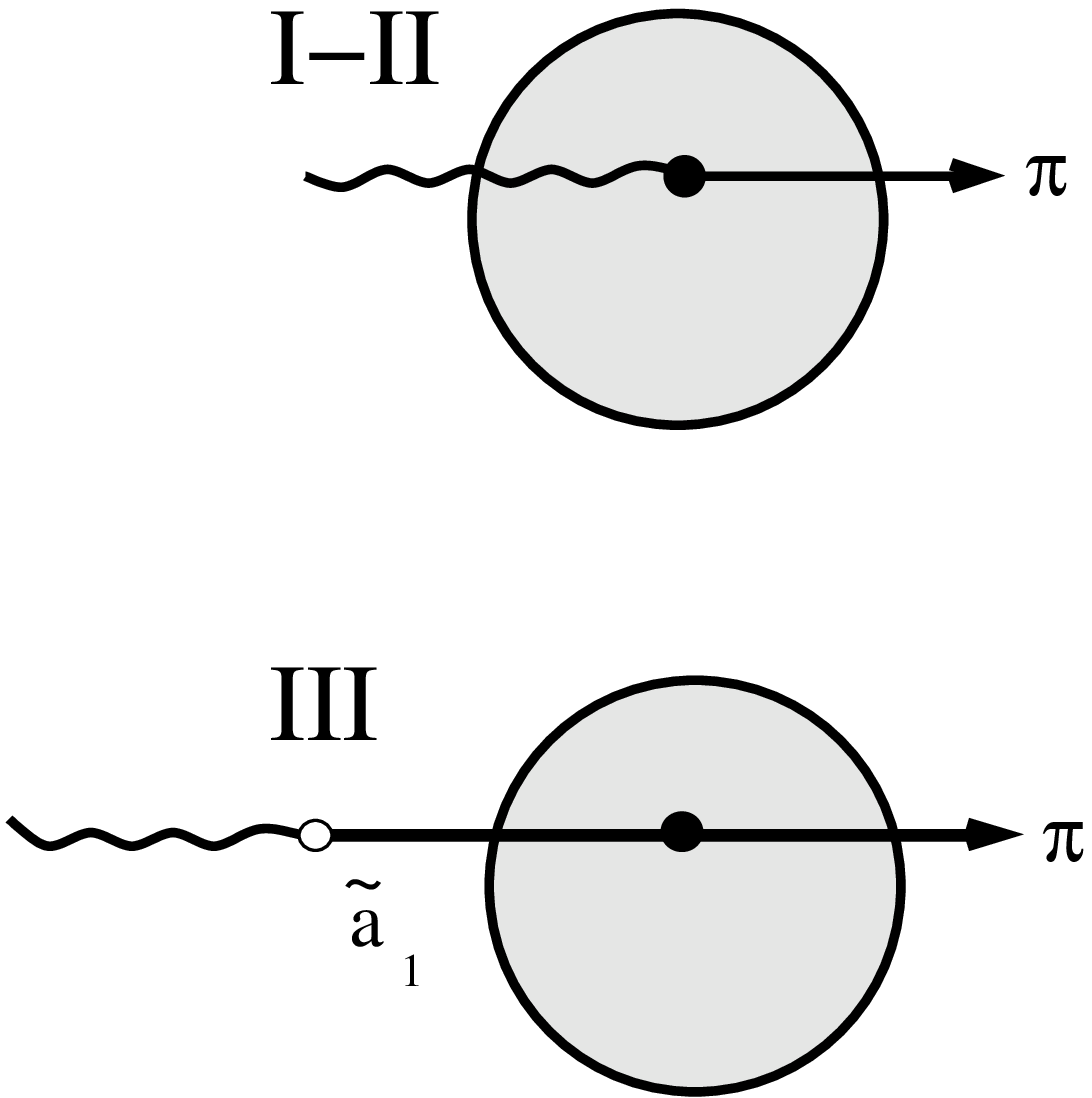}}}
\caption{{\it Left:} Invariant mass distribution of $3\pi$ produced diffractively as $1^+S$ wave in $\pi p\to\pi\rho p$. {\it Right:} two regimes for coherent pion neutrino-production off nuclei.
The upper and the lower figures correspond to the regimes described in the text, {\bf I-II} and {\bf III} respectively.
the waving line shows the axial current, $W$ or $Z$ bosons fluctuating to light hadronic states. }
\label{fig:pi-rho}      
\end{figure}

The problem is relaxed after inclusion of the $\rho\pi$ cut and other diffractive excitations into the dispersion relation
\cite{km,belkov}.   Indeed, the relation  
\beq
\sigma_{diff}(\pi p\to Xp)\approx \sigma_{el}(\pi p\to \pi p)
\label{140}
\eeq
does not contradict data.
The $\rho\pi$ cut can be represented by an effective 
pole $\tilde a_1$ \cite{belkov,asymmetry}, because the invariant mass distribution in the $1^+S$ wave in diffractive dissociation $\pi p\to 3\pi p$ indeed demonstrates a rather narrow peak shown in the left panel of Fig.~\ref{fig:pi-rho}. 
This bump is usually related to the non-resonance Deck mechanism \cite{deck}.
Thus, we arrive at the two-pole ($\pi+\tilde a_1$) model \cite{dispersion}.

\section{Absorptive corrections}\label{absorption}

\subsection{Incurable AR}\label{incurable}

Although the (approximate) equality (\ref{140}) seems to cure the striking controversy between the AR and data,
this cannot be an universal remedy for any target and at any energy. As an example, one can check the relation (\ref{140}) in the Froissart regime expected to onset at very high energies. In this regime $\sigma_{el}(\pi p\to \pi p)\to\sigma_{tot}/2$,
while $\sigma_{diff}(\pi p\to Xp)\propto \sigma_{tot}/\ln(s)$.
This happens because diffraction is suppressed by absorptive corrections,
while elastic cross section is enhanced. Thus, the condition~(\ref{140}) is badly broken in this regime. 

Another way to approach the black-disc regime of interaction is to use nuclear targets, for which the AR predicts
$\sigma^{\pi A}_{diff} \approx \sigma^{\pi A}_{el}$. However, such a relation should be severely broken because nuclear absorption enhances elastic, but suppresses diffractive interactions, so $\sigma^{\pi A}_{el}\propto A^{2/3}$,
but $\sigma^{\pi A}_{diff} \propto A^{1/3}$.

Thus, the AR is incurable:  even if the relation (\ref{140}) were valid for some target and at some energy, it will certainly fail for heavy nuclei, or/and at very high energies.
The diffractive diagonal and off-diagonal amplitudes cannot be universally related, since they are affected by absorptive corrections differently. 

\subsection{Time scales for diffractive neutrino-nucleus interactions}

In what follows we are going to test the AR in the process of diffractive pion production by neutrinos on proton, or nuclear targets, e.g. $\nu+p\to l+\pi+p$.
Within the considered two-channel model ($\pi+\tilde a_1$) the neutrino-nucleus interactions are characterized by two time scales: 
\beqn
t^\pi_c&=&\frac{2\,\nu}{m_\pi^2+Q^2}\\\nn
t^{\tilde a_1}_c&=&\frac{2\nu}{Q^2+m_{\tilde a_1}^2}.
\label{150}
\eeqn
The pion coherence time controls the interferences between 
pions produced at different longitudinal coordinates \cite{belkov,plb,jetp,gransasso}. The second time scale can be interpreted as the  $\tilde a_1$  fluctuation lifetime \cite{gransasso,dispersion}. 
Apparently, at small $Q^2\ll m_{\tilde a_1}^2$ these time scales are very different, $t^\pi_c\gg t^{\tilde a_1}_c$.
Depending on the transferred energy $\nu=E_\nu-E_l$ one can identify three different regimes of diffractive neutrino-nucleus interactions:

{\bf I.} $\nu < (Q^2+m_\pi^2)R_A$, which corresponds to rather low energies, approximately
$\nu\lesssim 500\MeV$.  In this regime the nuclear formfactor $F_A(q_L^2)$, where $q_L=1/t^\pi_c$, strongly suppresses the coherent production, which leaves the nucleus intact. Correspondingly, the cross section is steeply falling with the atomic number as $A^2\exp[-R_A^2/3(t^\pi_c)^2]$, where $R_A\propto A^{1/3}$ is the nuclear radius.

{\bf II.} $(Q^2+m_\pi^2)R_A < \nu < (Q^2+m_{\tilde a_1}^2)R_A$, what approximately corresponds to the energy range $0.5 \lesssim \nu \lesssim 40\GeV$. In this regime the absorptive corrections are caused by final-state interactions of the produced pion, while shadowing effects are still small. The cross section rises with $A$ as $A^{2/3}$.

{\bf III.} $\nu > (Q^2+m_{\tilde a_1}^2)R_A$, what approximately corresponds to energies $\nu > 40\GeV$.
This is the regime of maximal shadowing corrections \cite{morfin}. Neutrino fluctuates to a hadronic state, $\nu\to l+\tilde a_1$,
which diffractively produces a pion, $\tilde a_1+A\to \pi+A$. In this case only the periphery of the nucleus contributes, so the cross section is $\propto A^{1/3}$.

\begin{figure}[h!]
\centerline{
\resizebox{0.45\columnwidth}{!}{
\includegraphics{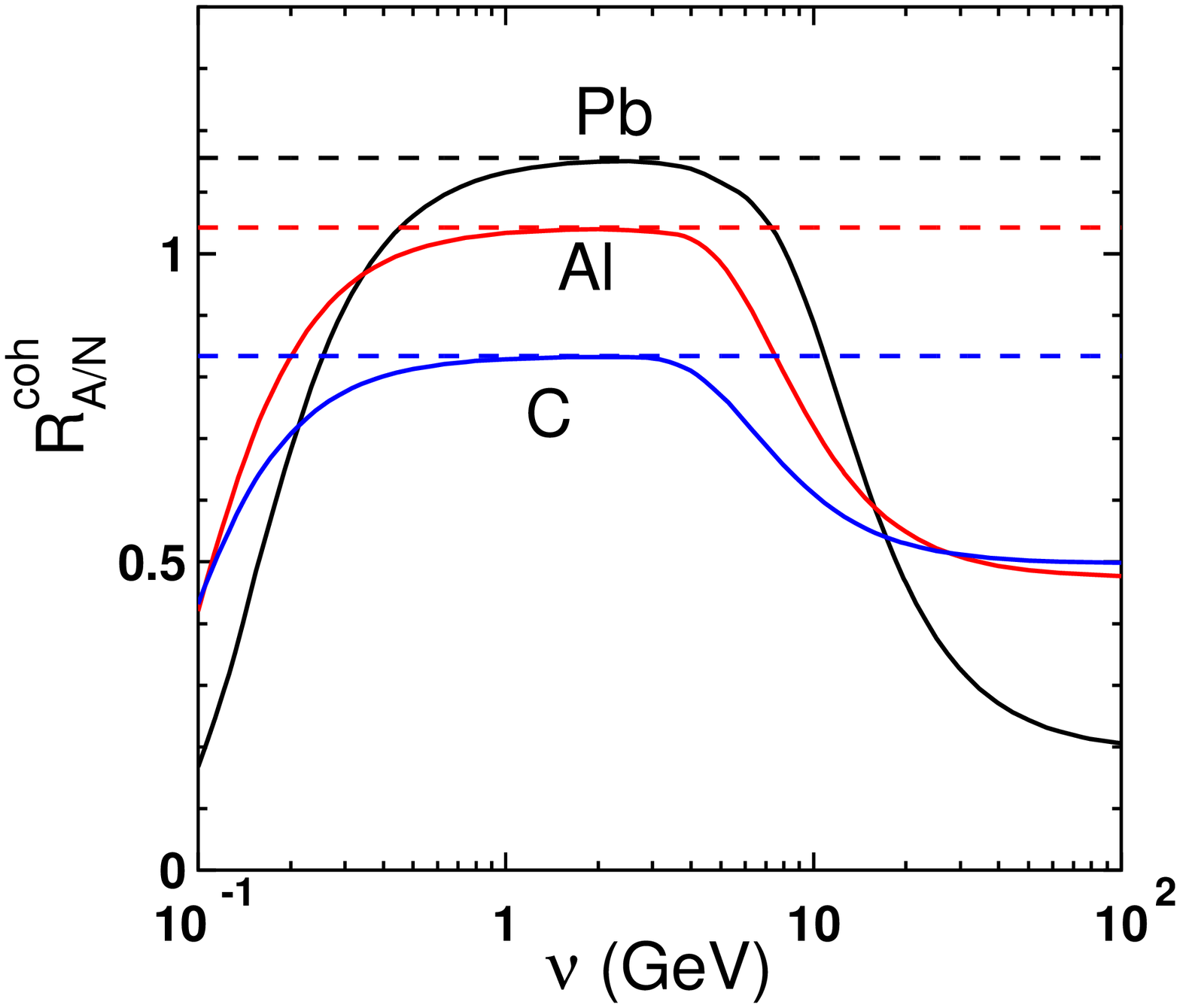} }
\hspace{0.5cm}
\resizebox{0.45\columnwidth}{!}{
\includegraphics{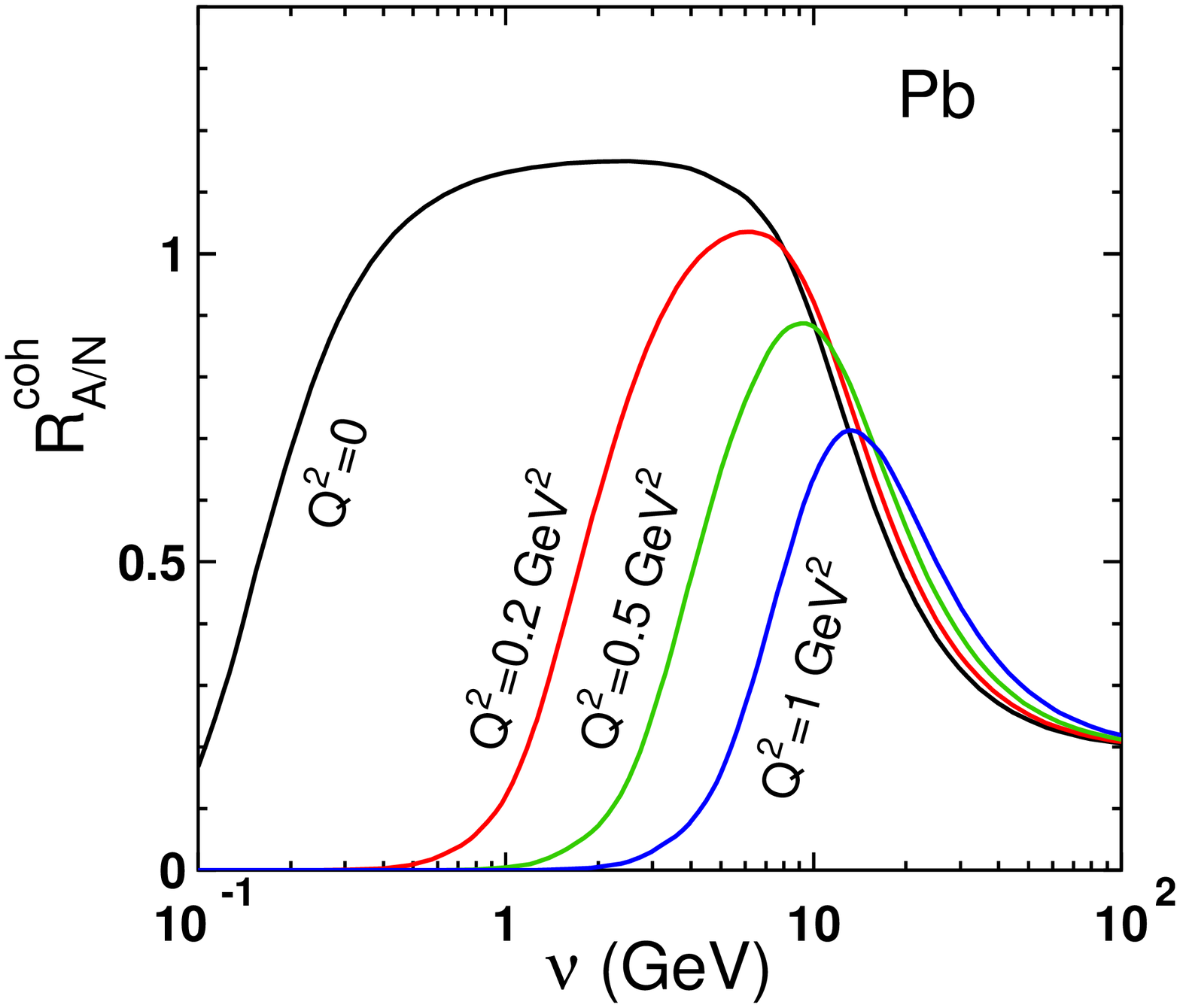}}}
\caption{{\it Left:} nuclear ratio Eq.~(\ref{160}) at $Q^2=0$ as function of the transferred energy $\nu$ for carbon, aluminum and lead. Dashed line show the expectations based on the AR. Solid curves show the results of calculations with the two-channel model \cite{dispersion}.
{\it Right:} the same as in the left figure, but only for lead and at different virtualities $Q^2$.}
\label{fig:Rcoh}      
\end{figure}

The nuclear effects for the cross section of coherent neutrino-production of pions are usually characterized by the ratio,
\beq
R^{coh}_{A/N}(\nu,Q^2) =
\frac{d\sigma(\nu A\to l\pi A)/dQ^2\,d\nu}
 {A\,d\sigma(\nu N\to l\pi N)/dQ^2\,d\nu}
\label{160}
\eeq
The results of calculations \cite{dispersion} within the two-channel model are shown in Fig.~\ref{fig:Rcoh}.
Comparison with the horizontal dashed lines, which present the expectations based on the AR, indeed demonstrates
existence of three energy regimes for the nuclear effects. At low energies, regime {\bf I}, the AR is trivially broken 
due to large longitudinal momentum transfer, which does not exist in pion-nucleus elastic scattering. In the regime {\bf II} the validity of the AR is surprisingly restored. The corresponding energy range is rather wide, $\nu$ varies by an order of magnitude. However at high energies in the regime {\bf III} the AR is severely broken by strong absorptive corrections. The cross section is below the AR prediction by factor $A^{1/3}$, which is a large number for heavy nuclei.

\subsection{Dipole representation}\label{dipole}

\begin{figure}[h!]
\centerline{
\resizebox{0.35\columnwidth}{!}{
\includegraphics{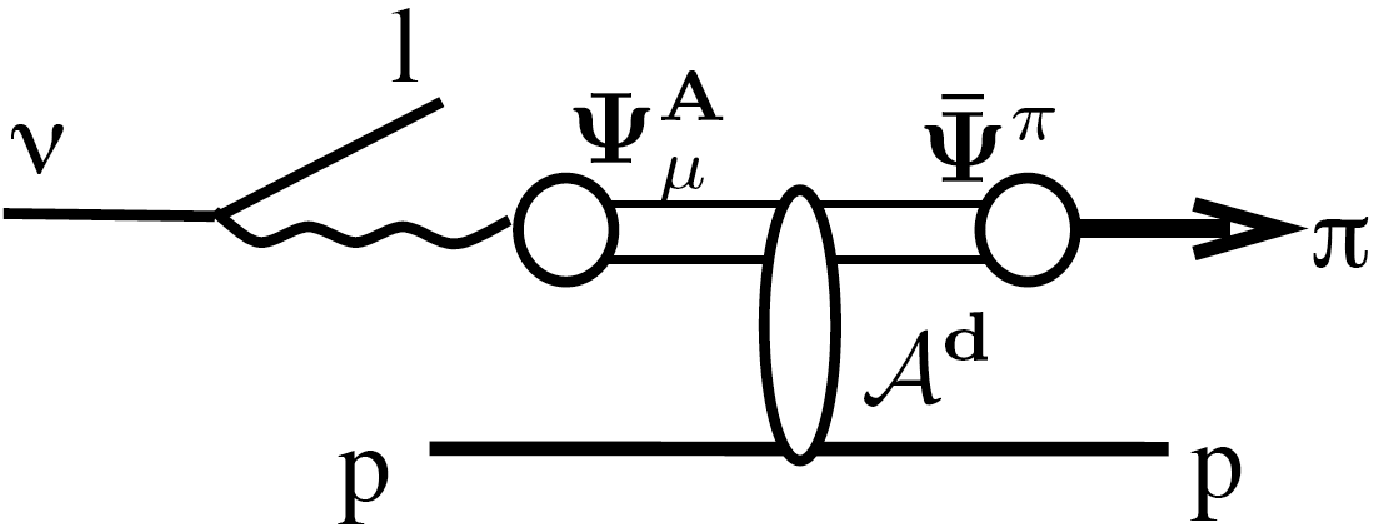} }
\hspace{1cm}
\resizebox{0.4\columnwidth}{!}{
\includegraphics{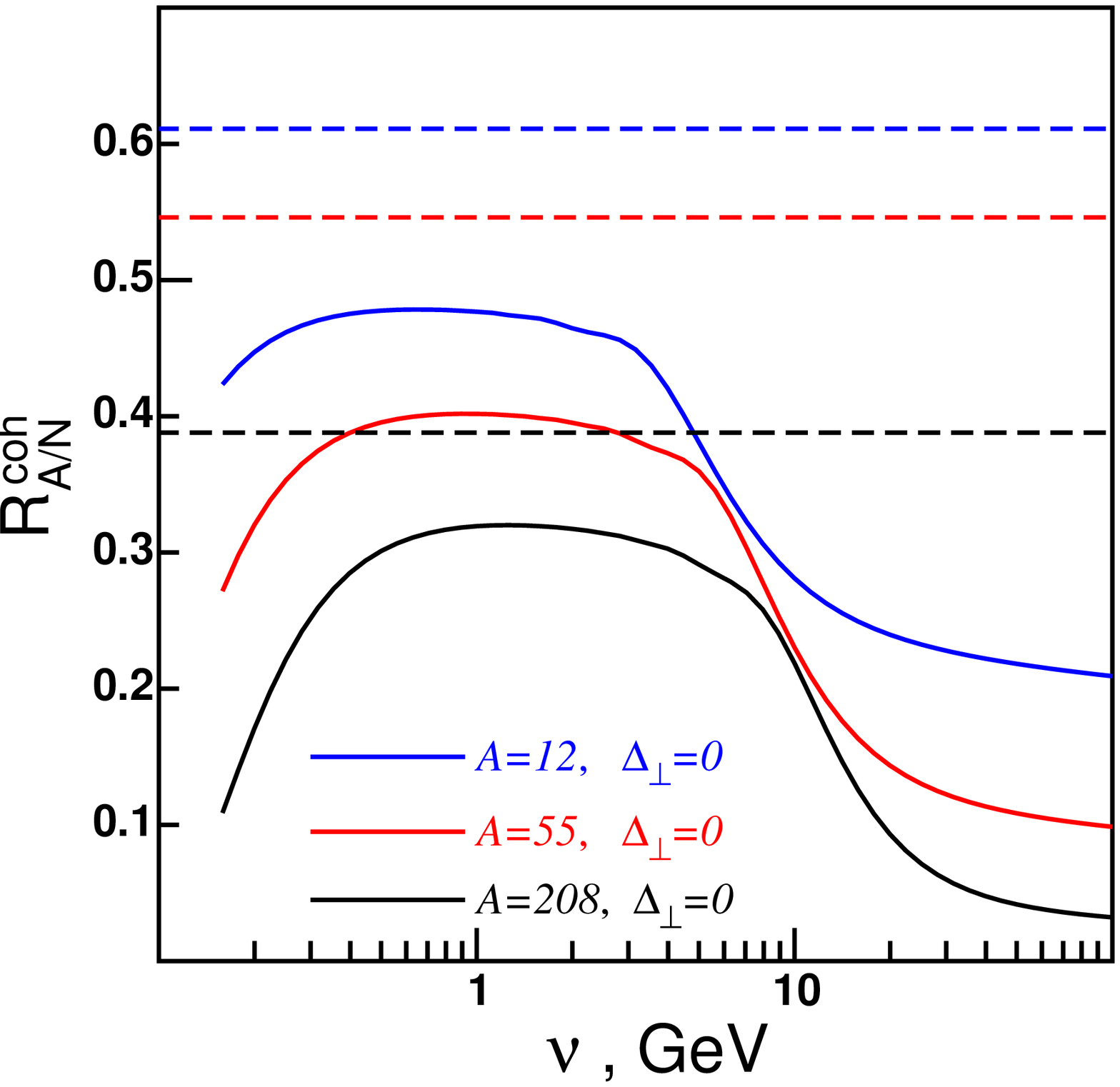}}}
\caption{{\it Left:} the hadronic fluctuation of the neutrino is a $\bar qq$ pair, which interacts elastically and then is projected into the pion light-cone wave function.
{\it Right:} The results of calculation within the dipole representation \cite{marat-A}, plotted by solid curves, are compared with the AR predictions, shown by horizontal dashed lines.}
\label{fig:Rcoh-dipole}      
\end{figure}

An extension of two channels to a multi-channel model would unavoidably lead to a large number of adjusted parameters. Instead, we switch to the dipole representation, which effectively includes all intermediate states in all orders \cite{zkl}. In this case pion production can be described as $\nu p\to l(\bar qq)p\to l\pi p$,
i.e. neutrino fluctuates into a $\bar qq$ dipole, which interacts elastically with the target and then is projected to the pion wave function \cite{marat-p}. This process is illustrated in the graph in the left panel of Fig.~\ref{fig:Rcoh-dipole}.
The cross section of this process has the form,
\beq
\nu\,\frac{d^{3}\sigma_{\nu p\to\mu\pi p}}{d\nu dtdQ^{2}}=
\frac{G_{F}^{2}\ L_{\mu\nu}\left(W_{\mu}^{A}\right)^{*}
W_{\nu}^{A}}{32\pi^{3}m_{N}^{2}E_{\nu}^{2}\sqrt{1+Q^{2}/\nu^{2}}},
\label{170}
\eeq
where
\beq
W_{\mu}^{A}\left(s,\Delta,Q^{2}\right) =
\int\limits _{0}^{1}d\beta d^{2}r\,\bar{\Psi}^{\pi}\left(\beta,{r}\right) 
\mathcal{A}^{d}\left(\beta,{r};\Delta\right)\Psi_{\mu}^{A}\left(\beta,{r}\right).
\label{180}
\eeq
Here $\mathcal{A}^{d}\left(\beta,{r};\Delta\right)$ is the dipole amplitude, which depends on dipole separation $r$, fractional light-cone momentum $\beta$ of $q$ or $\bar q$, and transverse momentum $\Delta$.
This amplitude is pretty well known from phenomenology, since it has been  fitted to photoproduction and DIS data.
The light-cone $\bar qq$ distribution amplitudes  $\Psi^A_\mu$ and $\Psi^\pi$   are calculated
in the instanton vacuum model \cite{ivm1,ivm2,ivm3} (see details in\cite{marat-p,marat-DA}).

The results for nuclear effects  are shown by solid curves in the right panel of Fig.~\ref{fig:Rcoh-dipole} in comparison with the AR predictions plotted by dashed lines. Differently from the results of the two-channel approximation, Fig.~\ref{fig:Rcoh}, now the cross section is quite below the AR benchmark at all energies, and it is easy to understand why. Within the two-channel approximation, the two time scales Eqs.~(\ref{150}) are well defined, and all events within the certain  energy range (see Fig.~\ref{fig:Rcoh}) belong to the regime {\bf II}, where the AR is valid.
On the contrary, the invariant mass of the dipole is not fixed, but varies with integration over $\beta$ and $r$.
The low mass tail of the dipoles can be associated with a long coherence time $t^{dip}_c\gg t^{\tilde a_1}_c$
which also gets into the regime {\bf III}, where the nuclear ratio is strongly suppressed (see Fig.~\ref{fig:Rcoh}).
As a result, the nuclear ratio plotted in Fig.~\ref{fig:Rcoh-dipole} is significantly below the level imposed by the AR.

\subsection{Incoherent diffractive neutrino-production of pions on nuclei}\label{incoh}

In a diffractive process the target nucleus can be excited and decay to fragments without new particle production.
Such a channel has a significant fractional cross section and is usually called incoherent production. In this case the nuclear ration is defined similar to (\ref{160}).
\beq
R^{incoh}_{A/N}(\nu,Q^2) =
\frac{d\sigma(\nu A\to l\pi A^*)/dQ^2\,d\nu}
 {A\,d\sigma(\nu N\to l\pi N)/dQ^2\,d\nu}.
\label{190}
\eeq

The nuclear effects for incoherent production $\nu+A\to l+\pi+A^*$ have different assignments for the regimes
considered above. The regime {\bf I} does not exist, because the nucleus breaks up and there is no nuclear formfactor
suppressing the cross section. So the regime {\bf II} starts at very low energies, and its atomic number dependence is $A^{2/3}$. This is much higher than the cross section calculated with the AR, which links this incoherent process to quasielastic pion-nucleus scattering. The latter is know to be significantly suppressed, as $A^{1/3}$
\cite{kps1}. Thus, the AR grossly underestimate the incoherent neutrino-production of pions at low energies.

\begin{figure}[h!]
\centerline{
\resizebox{0.45\columnwidth}{!}{
\includegraphics{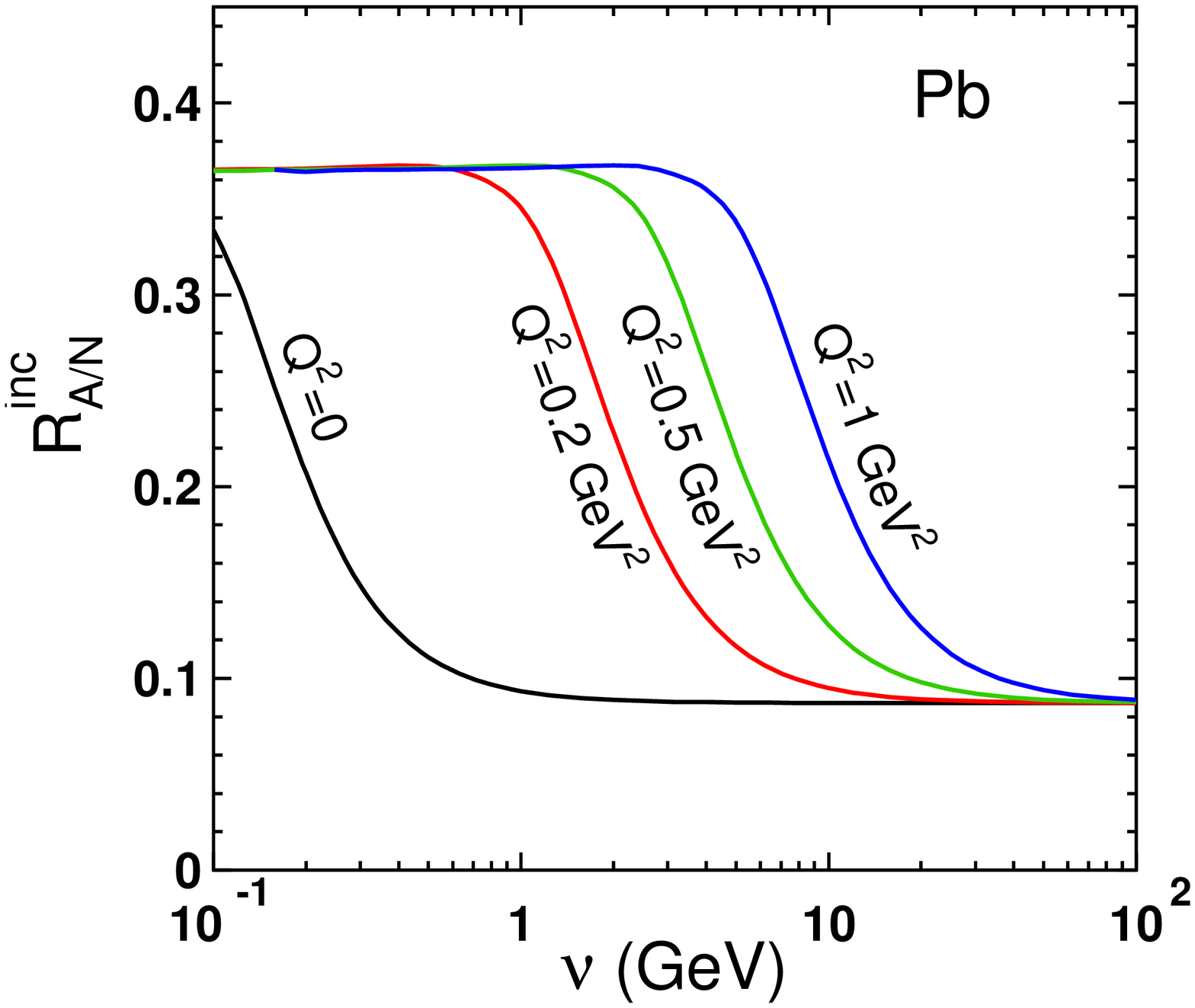} }
\hspace{0.5cm}
\resizebox{0.39\columnwidth}{!}{
\includegraphics{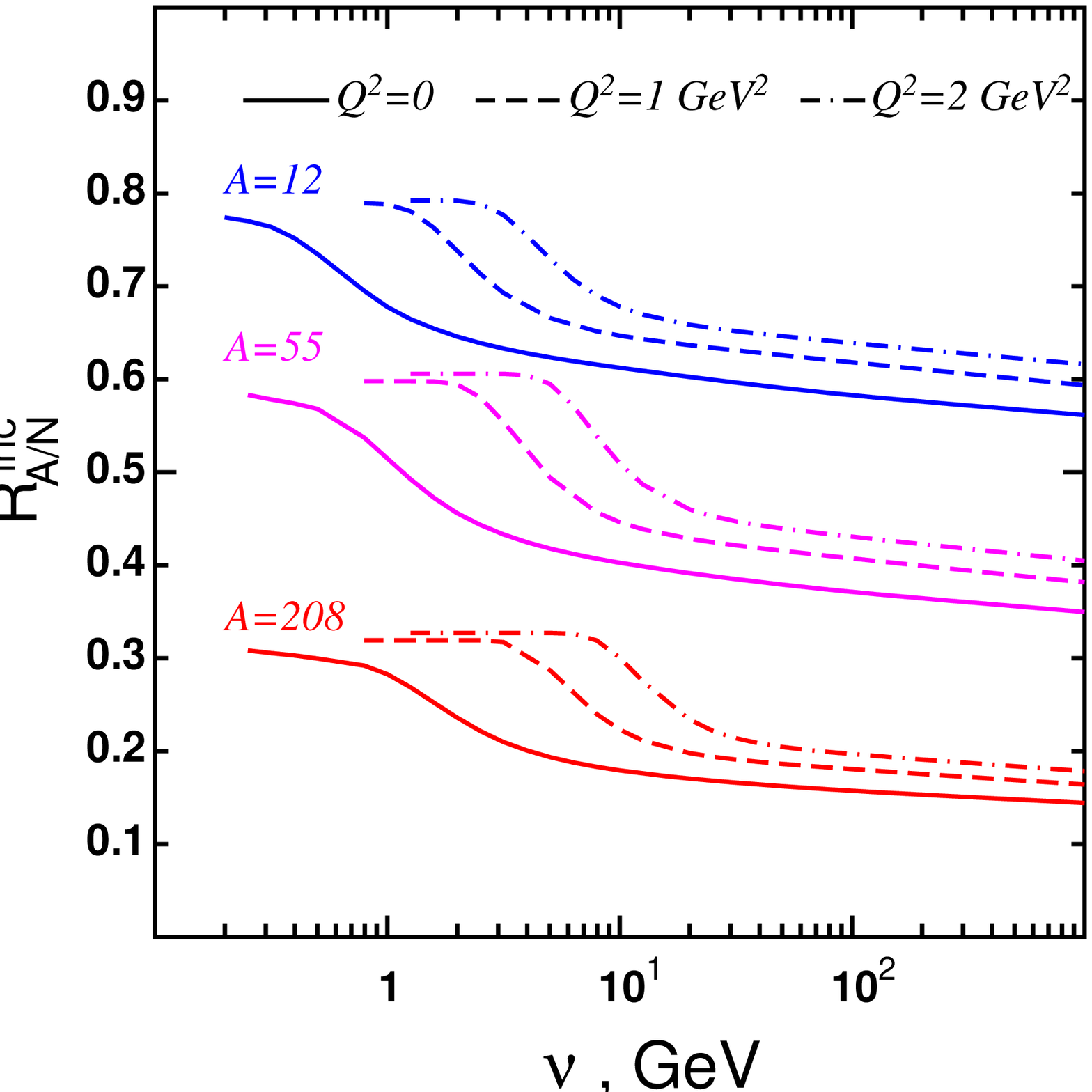}}}
\caption{The nuclear ratio Eq.~(\ref{190}) for incoherent neutrino-production of pions on nuclei as function of energy.
{\it Left:} The results of the two-channel approximation for lead at different virtualities $Q^2$.
{\it Right:}  The results of the dipole approach. }
\label{fig:Rincoh}      
\end{figure}

However at higher energies, in the regime {\bf III} the incoherent cross section drops to $A^{1/3}$, because both the $\tilde a_1$  and pion are subject to absorption,  and the two-channel model exactly reproduces the result of the AR \cite{dispersion}. The transition between the regimes {\bf II} and {\bf III} depends on $Q^2$ and is shown for lead in the left panel of Fig.~\ref{fig:Rincoh}.
We see that at $Q^2=0$ the AR is restored at $\nu>1\GeV$.

The dipole approach leads to similar results, much closer to the AR in the regime {\bf III} than for coherent production in the regime {\bf II}.
The reason is clear: in spite of the low-mass tail of the dipoles, which leads to a longer coherence time,
the process remains in the same regime {\bf III}. The results of the dipole approach 
are shown for several nuclei in the right panel of Fig.~\ref{fig:Rincoh}.

\section{Summary}\label{summary}

\begin{itemize}

\item
The Goldberger-Treiman relation is not a result of pion exchange, which is suppressed in $\beta$-decay and muon capture. This is a result of a miraculous link between light and heavy states.

\item
In the diffractive neutrino-production of pions PCAC establishes a link between diagonal and off-diagonal
amplitudes, which cannot be correct, because both are strongly and differently affected by absorption.

\item
The Adler relation for coherent neutrino-production of pions
is always broken, but especially at high energies. On the contrary, in incoherent pion production the Adler relation is broken at low, but is restored at high energies.

\end{itemize}

{\bf Acknowledgements:} B.Z.K. is thankful to the organizers of the International Conference on New Frontiers in Physics for the invitation to deliver  this talk. Our work was supported in part by Fondecyt (Chile)
grants 1090291, 1090236, 1100287 and 1120920.

\end{document}